%% file: hybridmb.tex
\documentclass[sigconf]{acmart}

\AtBeginDocument{%
  \providecommand\BibTeX{{%
    \normalfont B\kern-0.5em{\scshape i\kern-0.25em b}\kern-0.8em\TeX}}}

\setcopyright{rightsretained} 
\copyrightyear{2020} 
\acmYear{2020} 
\acmConference{SIGGRAPH '20 Posters}{August 17, 2020}{Virtual Event, USA}
\acmBooktitle{Special Interest Group on Computer Graphics and Interactive Techniques Conference Posters (SIGGRAPH '20 Posters), August 17, 2020}
\acmDOI{10.1145/3388770.3407436}
\acmISBN{978-1-4503-7973-1/20/08}

\acmSubmissionID{pos\_203}

\begin{document}

\title[Hybrid MBlur]{Hybrid MBlur: Using Ray Tracing to Solve the Partial Occlusion Artifacts in Real-Time Rendering of Motion Blur Effect}

\author{Tan Yu Wei}
\affiliation{%
 \institution{National University of Singapore}
}
\email{yuwei@u.nus.edu}

\author{Cui Xiaohan}
\affiliation{%
 \institution{National University of Singapore}
}
\email{cuixiaohan@u.nus.edu}

\author{Anand Bhojan}
\affiliation{%
 \institution{National University of Singapore}
}
\email{banand@comp.nus.edu.sg}

\renewcommand{\shortauthors}{Tan et al.}

\begin{abstract}
	For a foreground object in motion, details of its background which would otherwise be hidden are uncovered through its inner blur. This paper presents a novel hybrid motion blur rendering technique combining post-process image filtering and hardware-accelerated ray tracing. In each frame, we advance rays recursively into the scene to retrieve background information for inner blur regions and apply a post-process filtering pass on the ray-traced background and rasterized colour before compositing them together. Our approach achieves more accurate partial occlusion semi-transparencies for moving objects while maintaining interactive frame rates.
\end{abstract}



\begin{teaserfigure}
	\centering
	\includegraphics[width=\linewidth]{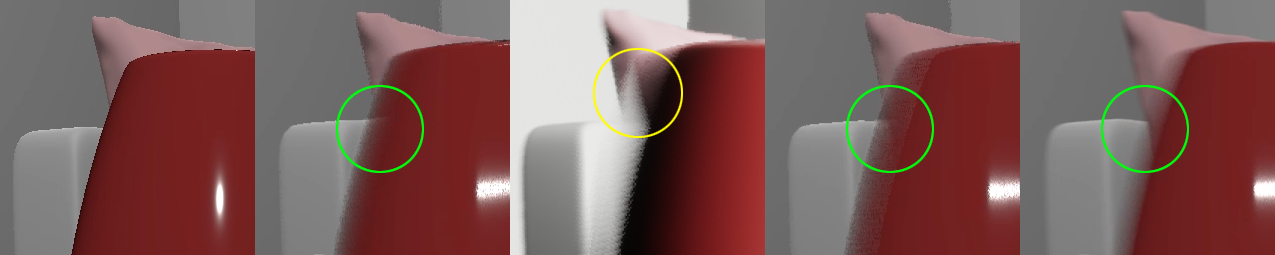}
	\caption{From left to right: original scene, adapted post-process \cite{McGuire:2012:RFP} (295 fps), UE4 post-process, hybrid (205 fps) and ray-traced MBlur on \href{https://www.blendswap.com/blends/view/75692}{\textsc{The Modern Living Room}} (\href{https://creativecommons.org/licenses/by/3.0/}{CC BY}) with a GeForce RTX 2080 Ti}
	\Description{Close-up shot of foreground vase whose left edge covers left edge of cushion in background}
	\label{fig:teaser}
\end{teaserfigure}

\maketitle
 
\input{introduction}
\input{design}
\input{discussion}

\begin{acks}
	This work is supported by the Singapore Ministry of Education Academic Research grant T1 251RES1812, “Dynamic Hybrid Real-time Rendering with Hardware Accelerated Ray-tracing and Rasterization for Interactive Applications”. 
\end{acks}

\bibliographystyle{ACM-Reference-Format}
\bibliography{hybridmb}

\end{document}

%% file: introduction.tex
\section{Introduction}

Blurred regions by nature uncover information of hidden background areas which are otherwise excluded from the render, as an effect of partial occlusion. Moving objects blur outwards and inwards at their silhouette, causing the region surrounding their silhouette to appear semi-transparent \cite{Jimenez:2014:ARR}. Outer blur represents an object's blur into its background, while inner blur refers to the blur produced within the silhouette of the object itself. 

One key to Motion Blur (MBlur) rendering is hence the recovery of background colour in inner blur regions which is inaccurate with screen space approaches. Post-process techniques like \citet{McGuire:2012:RFP} approximate the background of the inner blur with neighbouring pixels when the background colour of the target pixel cannot be retrieved from raster information. This approach not only produces a mere approximation of the true background geometry of inner blur regions, but also leads to inaccuracies between real and approximated backgrounds for sharp and blurred regions respectively. Our technique addresses these issues by obtaining the exact colour of occluded background with ray tracing for a more accurate MBlur.

%% file: design.tex
\section{Design}

We first obtain per-pixel information such as camera space depth, world space surface normal vector, screen space mesh ID and velocity as well as rasterized colour under deferred shading. The same depth and colour information for background geometry are retrieved by our novel ray reveal pass within a ray mask for pixels in the inner blur of moving foreground objects. A tile-dilate pass is then applied to these 2 sets of buffers to determine the sampling range of our gathering filter in the subsequent post-process pass which is adapted from \citet{McGuire:2012:RFP}. Both the ray-revealed result and rasterized output are then blurred by the post-process pass, and lastly composited together to produce our final image.

\subsection{Post-Process}

We gather samples from a heuristic range of nearby pixels and compute the amount of contribution of each sample \cite{McGuire:2012:RFP} to produce a motion-blurred effect separately for rasterized as well as ray-revealed information.

\begin{figure}[ht]
    \centering
    \includegraphics[width=\linewidth]{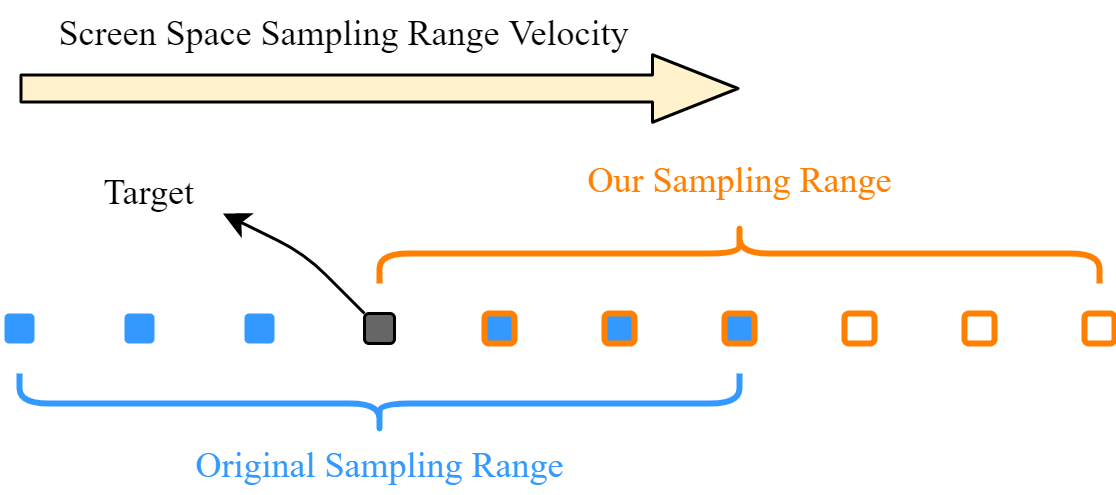}
    \caption{Range of samples for each pixel}
    \Description{Sampling range translated forward by half the magnitude of the velocity as compared to \citeauthor{McGuire:2012:RFP}}
    \label{fig:mb-samplerange}
\end{figure}

For the sampling range, \citeauthor{McGuire:2012:RFP} centers the sampling area at the pixel itself, creating a 2-directional blur effect towards both sides of the geometry edge as shown in \autoref{fig:mb-samplerange}. Although this produces more symmetric blur for thin objects and the specular highlight of curved surfaces, it introduces complexity in smoothening the transition between the inner and outer blur. Our approach refrains from this problem by letting the target pixel be at the end of the sampling area. In this way, inner and outer blur can be considered separately without much change to the pipeline.

For inner blur, \citeauthor{McGuire:2012:RFP} uses colour information of nearby samples based on the reasoning that approximate information, though inaccurate, is more desirable than the absence of any information whatsoever. However, this creates an inconsistency of background information at the silhouette of moving foreground objects. 

\subsection{Ray Mask}

We obtain true background information by shooting rays within a ray mask, where only inner blur regions corresponding to geometry edges are marked. For every pixel with a nonzero speed, we first translate it in the direction of its velocity for one magnitude of its estimated displacement within the exposure time and compare the change in depth value and mesh ID. For a pixel to be in the inner blur region of a foreground object, it would have to be shallower than its displaced position by a certain amount, and also correspond to a different mesh ID as part of a separate object. Pixels that do not satisfy these conditions are hence filtered out.

The remaining pixels are passed into a Sobel convolution kernel which detects edges based on the differences in surface normal and depth with surrounding pixels. The result is then evaluated against a high edge threshold to effectively eliminate non-edges from the edge mask. Lastly, a range check pass is applied, where each pixel with a nonzero speed samples along its estimated displacement. If any sample encountered is marked in the edge mask, the pixel passes into the final ray mask. 

\subsection{Ray Reveal}

We adapt recursive ray tracing and shoot rays within inner blur regions, iteratively advancing them deeper into the scene until a different object is found. When the recursion terminates, the background originally blocked at the current view is revealed using information from the final hit point. Although the accuracy of our motion-blurred background will increase when subsequent deeper layers are revealed with rays, for efficiency, we limit the ray reveal to one layer and apply post-process (i.e. approximate using neighbour information) to achieve our motion-blurred background.

At the pixel level, a ray is dispatched for each pixel within the ray mask and automatically advanced for recursion after the first hit. The advancing of a ray is performed by shooting a new ray along the same direction with its starting point at the latest hit. Currently, luminance is used as an indicator for identifying different objects. Hence, our approach terminates the recursive advancing process when the newest hit point reads a luminance different from the original hit, or the maximum recursion level is met. However, we hope to switch to using GeometryIndex in DirectX Raytracing Tier 1.1, which will enable the ray tracing shader to distinguish geometries. This will be more suitable than luminance in scenes with multiple overlapping objects of the same luminance value. 

%% file: discussion.tex
\section{Discussion}

We evaluate our method against the state-of-the-art post-process MBlur from UE4, which adopts a similar approach to \citeauthor{McGuire:2012:RFP}. UE4 also generates MBlur by comparing the velocity of each pixel to the aggregate pixel velocity in its neighbourhood.

As seen in \autoref{fig:teaser}, post-process MBlur incorrectly reconstructs the background by reusing neighbouring information available in the rasterized image. Background colour is hardly visible within the inner blur of the foreground vase for \citeauthor{McGuire:2012:RFP} as compared to UE4. However, it is clear that the background for UE4 is mirrored from that outside the inner blur as seen from the additional cushion edge annotated in yellow. For hybrid MBlur, we can observe the accurate cushion edge marked in green that corresponds well to ground truth distributed ray tracing. Here, we manage to sample what is lost from depth testing in the rasterization pass while maintaining interactive frame rates.

In future, we hope to better support nonlinear motion within the exposure time by extending our technique with curve sampling. We are also exploring the restriction of velocity computation to a pre-defined scene depth if geometry movement is localized.

%% file: hybridmb.bbl

\begin{thebibliography}{2}


\ifx \showCODEN    \undefined \def \showCODEN     #1{\unskip}     \fi
\ifx \showDOI      \undefined \def \showDOI       #1{#1}\fi
\ifx \showISBNx    \undefined \def \showISBNx     #1{\unskip}     \fi
\ifx \showISBNxiii \undefined \def \showISBNxiii  #1{\unskip}     \fi
\ifx \showISSN     \undefined \def \showISSN      #1{\unskip}     \fi
\ifx \showLCCN     \undefined \def \showLCCN      #1{\unskip}     \fi
\ifx \shownote     \undefined \def \shownote      #1{#1}          \fi
\ifx \showarticletitle \undefined \def \showarticletitle #1{#1}   \fi
\ifx \showURL      \undefined \def \showURL       {\relax}        \fi
\providecommand\bibfield[2]{#2}
\providecommand\bibinfo[2]{#2}
\providecommand\natexlab[1]{#1}
\providecommand\showeprint[2][]{arXiv:#2}

\bibitem[\protect\citeauthoryear{Jimenez}{Jimenez}{2014}]%
        {Jimenez:2014:ARR}
\bibfield{author}{\bibinfo{person}{Jorge Jimenez}.}
  \bibinfo{year}{2014}\natexlab{}.
\newblock \bibinfo{title}{Advances in Real-Time Rendering in Games, Part I:
  Next Generation Post Processing in Call of Duty: Advanced Warfare}.
\newblock
\newblock
\urldef\tempurl%
\url{http://advances.realtimerendering.com/s2014/\#_NEXT_GENERATION_POST}
\showURL{%
\tempurl}


\bibitem[\protect\citeauthoryear{McGuire, Hennessy, Bukowski, and
  Osman}{McGuire et~al\mbox{.}}{2012}]%
        {McGuire:2012:RFP}
\bibfield{author}{\bibinfo{person}{Morgan McGuire}, \bibinfo{person}{Padraic
  Hennessy}, \bibinfo{person}{Michael Bukowski}, {and} \bibinfo{person}{Brian
  Osman}.} \bibinfo{year}{2012}\natexlab{}.
\newblock \showarticletitle{A Reconstruction Filter for Plausible Motion Blur}.
  In \bibinfo{booktitle}{\emph{Proceedings of the ACM SIGGRAPH Symposium on
  Interactive 3D Graphics and Games}} (Costa Mesa, California)
  \emph{(\bibinfo{series}{I3D '12})}. \bibinfo{publisher}{ACM},
  \bibinfo{address}{New York, NY, USA}, \bibinfo{pages}{135--142}.
\newblock
\showISBNx{978-1-4503-1194-6}
\urldef\tempurl%
\url{https://doi.org/10.1145/2159616.2159639}
\showDOI{\tempurl}


\end{thebibliography}
